\documentclass[aps,twocolumn,superscriptaddress,floatfix,pra]{revtex4}

\usepackage{hyperref}
\usepackage{amsmath}
\usepackage{color}
\usepackage{graphicx}   

\newcommand{\sslash}{\mathbin{/\mkern-4mu/}}

\graphicspath{{figures/}}

\begin{document}

\title[]{Generating and Manipulating Quantized Vortices On-Demand in a Bose-Einstein Condensate: a Numerical Study}
\author{B.~Gertjerenken}
\affiliation{Department of Mathematics and Statistics, University of Massachusetts Amherst, Amherst, Massachusetts 01003-9305, USA}
\affiliation{Institut f\"ur Physik, Carl von Ossietzky Universit\"at, D-26111 Oldenburg, Germany}
\author{P.G.~Kevrekidis}
\affiliation{Department of Mathematics and Statistics, University of Massachusetts Amherst, Amherst, Massachusetts 01003-9305, USA}
\affiliation{Center for Nonlinear Studies and Theoretical Division, Los Alamos
National Laboratory, Los Alamos, NM 87544}
\author{R.~Carretero-Gonz{\'a}lez}
\affiliation{Nonlinear Dynamical Systems Group\footnote{{\tt URL:} http:$\sslash$nlds.sdsu.edu/},
Computational Science Research Center, and
Department of Mathematics and Statistics,
San Diego State University, San Diego, California 92182-7720, USA}
\author{B.P.~Anderson}
\affiliation{College of Optical Sciences,
University of Arizona, Tucson, AZ, 85721, USA.}


\date{\today}

\begin{abstract}
We numerically investigate an experimentally viable method, that we
will refer to as the ``chopsticks method'', for generating and manipulating on-demand several vortices in a highly oblate atomic Bose-Einstein condensate (BEC) in order to initialize complex vortex distributions for studies of vortex dynamics.
The method utilizes moving laser beams (the ``chopsticks'')
to generate, capture and transport vortices inside and outside the
BEC.
We examine in detail this methodology and show a wide parameter range of applicability
for the prototypical two-vortex case, and show
case examples of producing and manipulating several vortices for which there is no net circulation, equal numbers of positive and negative circulation vortices, and for which there is one net quantum of circulation.  We find that the presence of dissipation can help stabilize the pinning of the vortices on their respective laser beam pinning sites. Finally, we illustrate how to utilize laser beams as repositories
that hold large numbers of vortices and how to deposit individual
vortices in a sequential fashion in the repositories in order to construct superfluid flows about the repository beams with several quanta of circulation.

\end{abstract}


\maketitle


\section{Introduction}

The realm of atomic Bose-Einstein condensates (BECs)~\cite{review_dalfovo,becbook1,becbook2} has presented
a pristine setting where numerous features of the nonlinear dynamics of quantized vortices, vortex lattices, and other vortex distributions
can be both theoretically studied and experimentally observed.  Research in this domain has enabled observations of, for example, precession and excitation of few
vortices~\cite{And2000.PRL85.2857,Hal2001.PRL86.2922,Bre2003.PRL90.100403,NeelyEtAl10,dshall,dshall1,dshall3},
collective excitations and dynamics of vortex
lattices~\cite{Eng2002.PRL89.100403,Cod2003.PRL91.100402,Smi2004.PRL93.080406,Sch2004.PRL93.210403,Tun2006.PRL97.240402},
decay of multiply quantized vortices into singly quantized
vortices~\cite{S2Ket2,Iso2007.PRL99.200403},
decay of dark solitons into vortices and vortex
rings~\cite{And2001.PRL86.2926,Dut2001.Sci293.663,zwierlein_ring},
and generation of quantum turbulence~\cite{Hen2009.PRL103.045301,Nee2013.PRL111.235301}.  Additionally, experimental efforts directed towards new methods of vortex
detection~\cite{dshall,Wil2015.PRA91.023621} are motivated by the need for direct measurements of the dynamics of arbitrary distributions of vortices.  These, and numerous other
experiments~\cite{And2010.JLTP161.574}, demonstrate enormous progress towards developing a more complete understanding of vortex dynamics in BECs.

Nevertheless, to the best of our knowledge,
there has not been an experimental demonstration of a method to construct at will arbitrary (topological charge, as well as spatial position) distributions of more than two vortices in a BEC.  Such a method would enable detailed experimental studies of interactions of vortices with each other, with sound, and with trap impurities.  New methods to study the evolution of many-vortex states involving quantum turbulence, as well as chaotic vortex dynamics~\cite{vkouk1,vkouk2,GoodmanEtAl2014} would be available, and the role of dissipation in superfluid dynamics could be more precisely determined based on experimental data.
Here we extend a proof-of-principle experimental demonstration,
described in a companion article~\cite{CARLO},
of on-demand vortex generation and manipulation of two oppositely charged vortices to many-vortex distributions, and further characterize the experimental parameters that enable two-vortex generation and manipulation. We refer to this approach
as the ``chopsticks method''.
We consider a highly oblate harmonically trapped BEC that is pierced by multiple blue-detuned laser beams (that play the role of the chopsticks that manipulate
the vortices) whose positions and intensities can be dynamically controlled.
We examine conditions for which the motion of the laser beams nucleates vortices in such a way that individual vortices are pinned to distinct
laser beams during the nucleation process.  Our numerical results indicate that on-demand engineering of many-vortex distributions is an experimentally realistic possibility, and open up new directions for the study of vortex dynamics in BECs.

The prototypical case on which our numerical study is based involves
the presence of two blue-detuned Gaussian laser beams that pierce a
highly oblate BEC.  Each laser beam acts as barrier of relatively potential
energy $U_0$ that of the same order of magnitude than the BEC
chemical potential ---weak beams do not nucleate vortices.
Initially, the two beams are stationary and
overlap within the BEC.  Consider a case where both beams begin
to move in the $y$ direction at a velocity that is significantly
lower than the critical speed for vortex {\em dipole} nucleation.
As described in a companion paper~\cite{CARLO}, the
beams push atomic superfluid out of the way, and superfluid fills in the space vacated by the laser beam.  Simultaneously, the beams are pulled apart in the $x$ direction, each beam having equal but opposite $x$-velocity component. Then, although the beams are always travelling at a speed well below the critical speed for vortex dipole nucleation, the ``holes'' (and the specific path they take) created by the laser beams facilitates the formation of two singly quantized vortices of opposite circulation that are simultaneously created and pinned (one per beam). Further slow, adiabatic, motion (below the critical speed)
of the laser beams, which serve as vortex optical tweezers,
guides the position of the pinned vortices.  After using the
chopstick beams to transport the vortices to desired locations, they can be released into the BEC to evolve freely by ramping off the laser beams.  It is worthwhile to add here that recently an {\it attractive} impurity
has been utilized in a one-dimensional setting, in order to perform
similar manipulations for one or more dark solitons~\cite{jans}.

Here we numerically focus on the use of several laser beams
to create and manipulate several vortices within the BEC.
We show that by moving the laser beams to desired locations and then ramping them off,
releases the vortices into the BEC as they subsequently evolve according to their
inter-vortex dynamics.   Since the basic vortex generation process creates two vortices of opposite circulation,  neutral vortex charge configurations with equal numbers of vortices of positive and negative circulation can be readily created. However, driving some of the
chopsticks beams out of the condensate, non-neutral charge configurations can also be generated.  Furthermore, by driving multiple chopsticks beams together or onto a separate ``repository'' laser beam, multiply quantized circulations about a single laser beam can be generated
and stored.
We confirm that by using
either many moving laser beams or by depositing vortices to stationary repository beams
and re-initiating the process of creating two pinned vortices with two chopsticks beams,
arbitrary amounts and configurations of
quantized vorticity can be prepared in the system, enabling the
examination of a wide array of associated phenomena.
An experimental realization of the prototypical scenario considered
herein involving two beams is explored in the companion paper~\cite{CARLO}.

Our discussion is organized as follows.
In Sec.~\ref{sec:model}, we briefly discuss the theoretical model that is used for
our study, namely the two-dimensional Gross-Pitaevskii equation
in the presence of a parabolic trap and a set of localized movable
defects. In Sec.~\ref{sec:num}, we present our numerical results.
We start with the simplest case of creating two vortices with two
chopsticks beams, which constitutes our
benchmark for quantifying the success of the method, and describe how the vortex generation and trapping process depends on the beam parameters.    Subsequently, we demonstrate that the process can be scaled up to
neutral configurations of more than two vortices.  We then consider the removal of a single vortex from neutral configurations by removing one
chopstick beam.  For the case of two initial vortices, this allows a single vortex to remain in the condensate at a location that is determined by the remaining beam. More generally, removing a single vortex by driving a beam out of the condensate leaves an imbalanced vortex charge configuration.
Although in the present work, we will consider imbalances leading to
a total charge of $\pm 1$, it will be evident that the method
can enable arbitrary such imbalances in the system.
Finally, we explore a number of variants
to the problem.
To distill out acoustic energy from these latter vortex configurations, we explore the effect of
thermally-induced dissipation~\cite{QuantumGases}, analyzed for vortices, e.g.,
in Ref~\cite{stoof,RooneyEtAl13,YanEtAl2014} (see also references therein).
We also present the possibility of
depositing large amounts of vorticity of the same
charge in a vortex repository.
In many cases, we explore the vortex dynamics and distributions that result from ramping off the laser beams.
Finally, in Sec.~\ref{sec:conclu}, we summarize
our findings and discuss directions for future study.


\section{Model}
\label{sec:model}

We numerically investigate a Bose-Einstein condensate in the presence of a strongly anisotropic trapping potential~$V_{\rm{ext}}=\frac{1}{2}m\left(\omega_x^2x^2+\omega_y^2y^2+\omega_z^2z^2  \right)$ with $\omega_x=\omega_y\equiv\omega_{\perp}\ll \omega_z$. In this case the trapped BEC acquires a nearly planar ``pancake'' shape. We start from the three dimensional Gross-Pitaevskii equation
\begin{equation}
i\hbar\partial_t\Psi = -\frac{\hbar^2}{2m}\Delta \Psi + g_{3\rm{D}}|\Psi|^2\Psi + V_{\rm{ext}} \Psi
\end{equation}
with $g_{3\rm{D}} = \frac{4\pi\hbar^2a_{s}}{m}$, where~$a_{s}$ denotes the $s$-wave scattering length. The wave function can be factorized into $\Psi = \Phi(z)\psi(x,y,t)$, where $\Phi(z)$ is the ground state of the respective harmonic oscillator: $\Phi(z)=\left(\frac{m\omega_z}{\pi\hbar}\right)^{1/4}\,{e}^{-m\omega_z\frac{z^2}{2\hbar}}$ and $\psi(x,y,t)$ is normalized to the number of atoms. Multiplying Eq.~(1) by $\Phi^*(z)$ and integrating over all $z$ yields the 2D GPE:
\begin{equation}
i\hbar\partial_t\psi = -\frac{\hbar^2}{2m}\Delta \psi + g_{2\rm{D}}|\psi|^2\psi + V_{\rm{ext}}(x,y,z=0) \psi.
\end{equation}
Here, we have introduced the 2D interaction parameter $g_{2\rm{D}} = g_{3\rm{D}}/(\sqrt{2\pi}a_{z})$ and the harmonic oscillator length $a_{z}=\sqrt{\hbar/m\omega_z}$.

We non-dimensionalize the 2D GPE by setting $\tilde{t} = t\omega_z$, $(\tilde{x},\tilde{y})=(x,y)/a_z$,  and $\tilde{\psi} = a_z \psi$.   The dimensionless 2D GPE then becomes
\begin{equation}
i\tilde{\psi}_{\tilde{t}} + \frac{1}{2}\left(\tilde{\psi}_{\tilde{x}\tilde{x}}+\tilde{\psi}_{\tilde{y}\tilde{y}} \right) - g|\tilde{\psi}|^2 \tilde{\psi}- V\, \tilde{\psi} = 0,
\end{equation}
where $g = 4\pi a_{s}/(\sqrt{2\pi}a_{z})$
and $V=\frac{\Omega^2}{2}(\tilde{x}^2+\tilde{y}^2)$,
where the effective (2D) harmonic trapping is $\Omega = \omega_{\perp}/\omega_z$.
Further letting $u = \sqrt{g}\tilde{\psi}$,
and dropping all tildes hereafter for notational simplification,
the 2D GPE for the case of repulsive interatomic interactions then becomes
\begin{equation}
iu_t + \frac{1}{2}\left(u_{xx}+u_{yy} \right) - |u|^2- V\, u = 0.
\end{equation}

In the presence of $N$ Gaussian laser beams of identical $1/e^2$ radii $\sigma$ (a dimensionless length, measured in units of $a_z$), the external effective (2D) potential is given by
the following combination of a harmonic trapping and the laser beams:
\begin{equation}
V(x,y,t)=\frac{\Omega^2}{2}(x^2+y^2)
+ \sum_{j=1}^N U_{0,j} \,{e}^{-2 \frac{x^2_j(t)+y^2_j(t)}{\sigma^2}},
\end{equation}
where, for the $j${th} laser beam, $U_{0,j}$ and $(x_j(t),y_j(t))$ are, respectively,
the time-dependent height of the light-induced barrier measured in units of $\hbar\omega_z$
and its position measured from the trap center in units of $a_z$.

For ease of evaluation of the experimental possibility of utilizing our vortex generation and manipulation methods in the discussion below, we define and utilize a \emph{dimensional} measure of the full width at half-maximum of the chopsticks beams as $b_{w} = a_z \sqrt{2\ln(2)}\sigma$.   The beam height $U_{0,j}$ for each beam is given in terms of the chemical potential
$\mu$ measured in units of $\hbar\omega_z$.
Furthermore, we specify all velocities in terms of the maximum sound speed
\begin{equation}
\label{eq:sound}
c = \frac{2\hbar}{m}\sqrt{\pi a_{s}n_{\max}},
\end{equation}
where $n_{\max}$ is the maximum of the three-dimensional atomic density
(usually attained at the center of the parabolic trapping)
\begin{equation}
n_{\max} = \left(\frac{m\omega_z}{\pi\hbar}\right)^{1/2}\,
\max(|u|^2)/\left(\frac{4\pi a_{s}a_{z}}{\sqrt{2\pi}}\right).
\end{equation}

Our procedure is conceptually similar to that of the
companion article~\cite{CARLO}, where blue detuned
laser beams are used in order to produce repulsive barriers
that generate and then dynamically manipulate vortices inside the BEC.  We specify parameters used in the calculations, and present the results of our numerical simulations using those parameters.

As mentioned previously, in recent years a significant consideration
in connection to vortices concerns their dynamics in the presence
of thermally induced dissipation; see, e.g.,
Refs.~\cite{QuantumGases,stoof,RooneyEtAl13,YanEtAl2014}
for which the system is described by the
dissipative Gross-Pitaevskii equation (dGPE)
\begin{equation}\label{eq:diss}
(i-\gamma)u_t + \frac{1}{2}\left(u_{xx}+u_{yy} \right) - |u|^2 u - V\, u = 0,
\end{equation}
where $\gamma$ is a dimensionless damping constant.
As we show below, considering dissipation in our system, not only corresponds
to a more realistic experimental scenario, but it helps to remove vortices
from the periphery of the BEC cloud as dissipation induces them to spiral outwards.

\section{Numerical results}
\label{sec:num}

The initial condition for our evolutionary dynamics is obtained by
a 2D fixed point iteration (a Newton's method) in order
to identify the ground state of the system in the presence of an
even number of beams, ranging from two to eight.
This state is devoid of vortices. Subsequently, we
compute the time-evolution
using a variable-order Adams PECE algorithm, of the type originally
elaborated in Ref.~\cite{ShampineGordon75}.

Although our findings can be straightforwardly
generalized to different trapping and
atomic gas parameters, for concreteness within our pancake-shaped geometry,
we choose parameter values consonant with the experiments of
the companion article~\cite{CARLO}.
Namely, we choose $\omega_{\perp} = 2\pi\times 2$~Hz, $\omega_z = 2\pi\times 90$~Hz,
and $a_{s} = 5.3$~nm corresponding to $^{87}$Rb.  We examine a range of chemical potentials, and indicate this value of $\mu$ for each case study.  These parameters correspond to dimensionless times measured in units of 1.77 ms.

\begin{figure}
\begin{center}
\includegraphics[width = 1.0\linewidth]{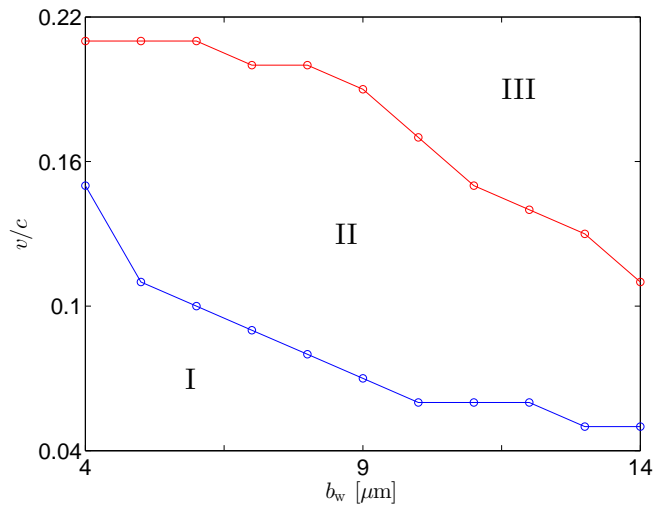}
\end{center}
\caption{(Color online) Regions of successful and unsuccessful vortex generation and trapping
as a function of the beam width~$b_{w}$ and the scaled chopsticks speed~$v/c$. The chemical potential corresponds to~$\mu=1.5$ and the beam height is~$2\mu$. Both beams are initially situated at $(0,-20)$~$\mu$m and moved diagonally on straight lines with the corresponding velocity to their final positions at $(\pm 10,0)$~$\mu$m.
Region~I: The velocity of the chopsticks is too small and vortices are not created.
Region~II: Two vortices of opposite charge are created, successfully trapped and dragged along with the laser beams.
Region~III: The velocity of the laser beams is too large, resulting in the vortices being lost from their respective beams and remaining behind the beams.
For beam widths approximately smaller than $b_{w} = 4$~$\mu$m no vortex trapping is supported:
any vortices created are immediately lost from the beams.
Movies showing the full time evolution exemplary for each of the three regimes
(and for each of the following figures) can be found at
\href{http://nonlinear.sdsu.edu/~carreter/Chopsticks.html}%
{http:$\sslash$nonlinear.sdsu.edu/$\sim$carreter/Chopsticks.html}
}
\label{fig:Fig1}
\end{figure}

\subsection{Neutral vortex configurations}

First we discuss the generic example of the creation and trapping of a pair of oppositely charged vortices that can later be used as a building block for the generation of a larger distribution of an even number of vortices, with equal numbers of positively and negatively charged trapped vortices. Two beams of equal waist and height are initially
located at the same position and are subsequently
moved with the same speed under a suitable angle to two different final
positions. For a given value of the beam size $b_{w}$, there is a range of suitable values of the beam velocity, exceeding a (lower) critical
value but much less than the speed of sound, for which the flow around the beams results in the generation and pinning of two vortices with opposite charge.  If the velocity is chosen to be higher than this range, vortices are created but cannot remain trapped by their respective beams.  These vortices start lagging
behind their respective beams and finally are released from the trapping action of the chopsticks beams.  The vortices may then annihilate one another.
Given that this is a prototypical {\it benchmark} scenario, in
Fig.~\ref{fig:Fig1} we display regions of successful and unsuccessful vortex generation and stable trapping and manipulation, for various beam widths~$b_{w}$ and
beam speed~$v/c$ [measured in units of the maximum speed of sound; see Eq.~(\ref{eq:sound})] for an experimentally realistic
chemical potential~$\mu=1.5$, which corresponds to $\sim 4.4 \times 10^5$ atoms comprising the BEC for the parameters given previously, a radial Thomas-Fermi BEC radius of $\sim 87\,\mu$m, and a BEC healing length of $\sim 0.64 \,\mu$m. The waist of the beams should be
large enough to support the existence and trapping of the vortices.
We find that for the BEC parameters used in our study, with $\mu \sim 1.5$,
the beam width approximately has to exceed~$4$~$\mu$m. For beam widths below
this value, the vortices created are immediately expelled from the
trapping beams~\cite{kody}.
We have ued throughout this study a beam height twice as
large as the BEC chemical pontental ($U_{0,j}=2\mu$).
However, we have checked that for lower beam heights between
$U_{0,j} = \mu$ and $U_{0,j} = 2\mu$, vortex nucleation is
still successful for beam widths approximately larger than~$4$~$\mu$m.

\begin{figure}
\begin{center}
\includegraphics[width = 1.0\linewidth]{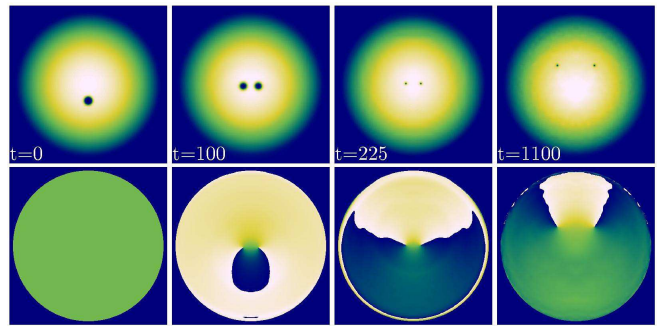}
\end{center}
\caption{(Color online) Controlled generation of a vortex dipole for $\mu=1.5$.
Throughout this manuscript the upper row shows the atomic density in false color
(red corresponds to highest densities, blue to lowest) while the lower row
displays the phase (blue to red corresponds to phases from 0 to $2\pi$) at the
times indicated in the legend.
A pair of beams with $b_{w} = 6$~$\mu$m and beam height $U_{0,j}=2\mu$ initially ($t=0$) placed
at $(0,-20)$~$\mu$m are moved within~$\Delta t = 110$ to $(\pm 10,0)$~$\mu$m. The process
nucleates a vortex dipole that is dragged along to the desired location.
From $t=110$ to $t=210$ the beams are linearly ramped down to release the vortices
to the free time evolution where the vortices start the typical motion of a symmetric
dipole pair.
The numerics for $\mu=1.5$ are run on $749\times 749$ grid points. The resulting
spatial discretization~$dx=0.23$ guaranties the existence of about 10 grid points
within the vortex profile.
}
\label{fig:N2sym}
\end{figure}

\begin{figure}
\begin{center}
\includegraphics[width = 1.0\linewidth]{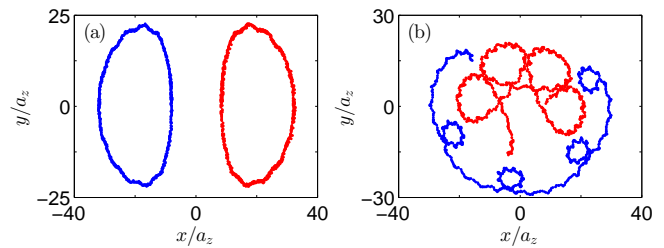}
\end{center}
\caption{(Color online)
(a) Vortex trajectories for the symmetric vortex configuration from Fig.~\ref{fig:N2sym} displayed from the release of vortices at $t=210$ up to~$t=4000$.
(b) Vortex trajectories for the asymmetric configuration of Fig.~\ref{fig:N2asym} from release
of vortices up to~$t=16000$.}
\label{fig:N2traj}
\end{figure}

\begin{figure}
\begin{center}
\includegraphics[width = 1.0\linewidth]{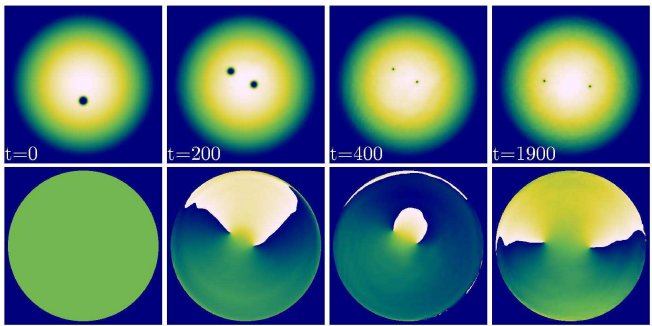}
\end{center}
\caption{(Color online)
Similar to Fig.~\ref{fig:N2sym} but for an asymmetric vortex configuration.
While one beam is kept fixed after $t=110$ at $(10,0)$~$\mu$m, the other beam
is moved further until $t=200$. Then both beams are kept fixed in this asymmetric
configuration and are linearly ramped down within $\Delta t= 100$.
As a result, asymmetric vortex dipole dynamics arises past the
ramp-down time.
}
\label{fig:N2asym}
\end{figure}

We now examine the trajectories of two oppositely charged vortices that result
from turning off the chopstick beams, in the case of
typical symmetric and asymmetric configurations of the vortices~\cite{dshall1}.
We choose a value of~$\mu=1.5$ for the chemical potential,
a beam waist~$b_{w} = 6$~$\mu$m and a beam height~$U_{0;j}=2\mu$.
It is worth mentioning at this stage that lower values of the chemical potential resulted
in large (density modulation) disturbances  as the chopsticks move through the
condensate, especially if more than two beams
are present; on the other hand, the case of larger values of the chemical potential is more difficult
to track numerically as the vortex width becomes relatively small compared to the
numerical domain and thus a  large numerical grid is necessary.
Therefore, for larger numbers of trapped vortices, discussed later,
it is necessary to increase the value of~$\mu$ at the expense of more intensive
numerics. The typical generation of a
symmetric vortex configuration is shown in Fig.~\ref{fig:N2sym}.
After the vortices have been created and have reached the desired symmetric
final positions $(\pm 10,0)$~$\mu$m the beams are
adiabatically and linearly ramped down to release the vortices to undergo free
(i.e., uninhibited by the presence of the chopsticks beams)
time evolution. As a result, the released vortices
exhibit the typical dynamical features of a 
vortex dipole configuration~\cite{dshall1,dshall4}. 
This is illustrated in Fig.~\ref{fig:N2traj}(a) for the
symmetric case of Fig.~\ref{fig:N2sym}.
Figure~\ref{fig:N2asym} shows a similar example as the one depicted in
Fig.~\ref{fig:N2sym} but for an asymmetric motion of the chopsticks.
In this case, one laser beam is kept fixed after $t=110$ at $(10,0)$~$\mu$m
while the other beam is moved further until $t=200$ where both beams are kept
fixed and ramped down. This procedure seeds an asymmetric configuration
that, after removal of the chopsticks, evolves in the typical
epitrochoidal trajectories for asymmetric vortex dipoles~\cite{dshall1,dshall4}
as shown in Fig~\ref{fig:N2traj}(b).

\begin{figure}
\begin{center}
\includegraphics[width = 1.0\linewidth]{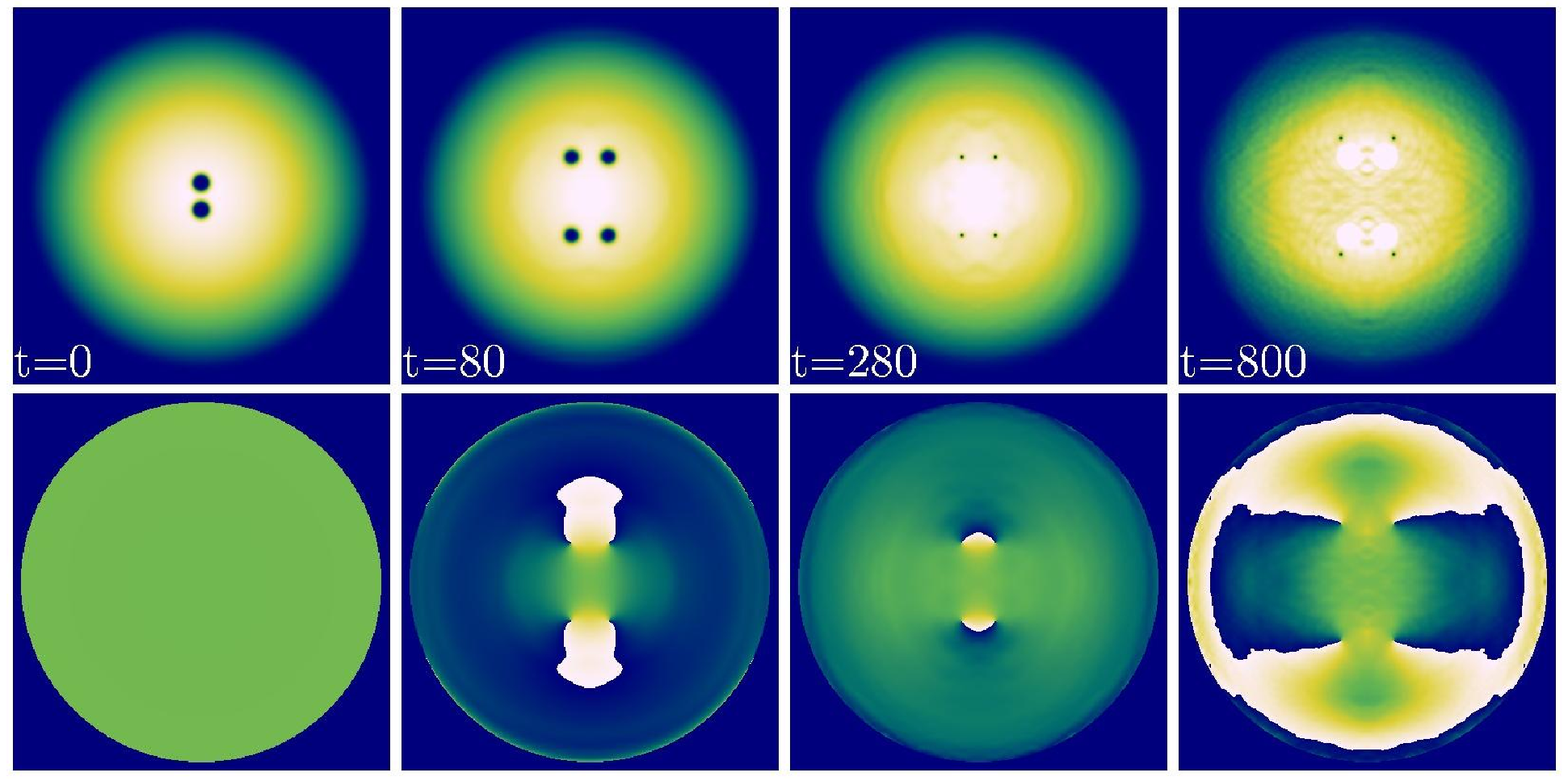}
\includegraphics[width = 1.0\linewidth]{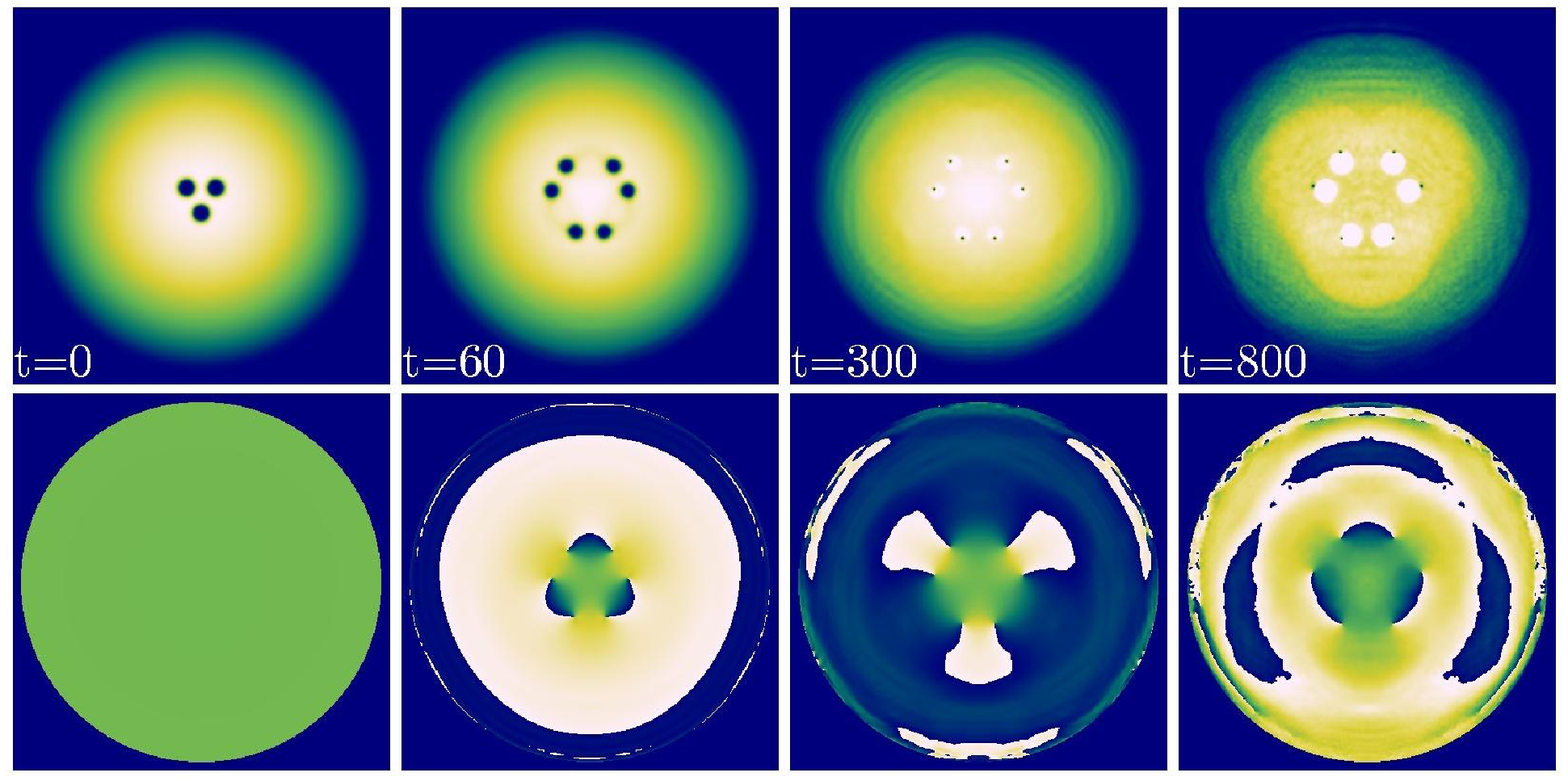}
\includegraphics[width = 1.0\linewidth]{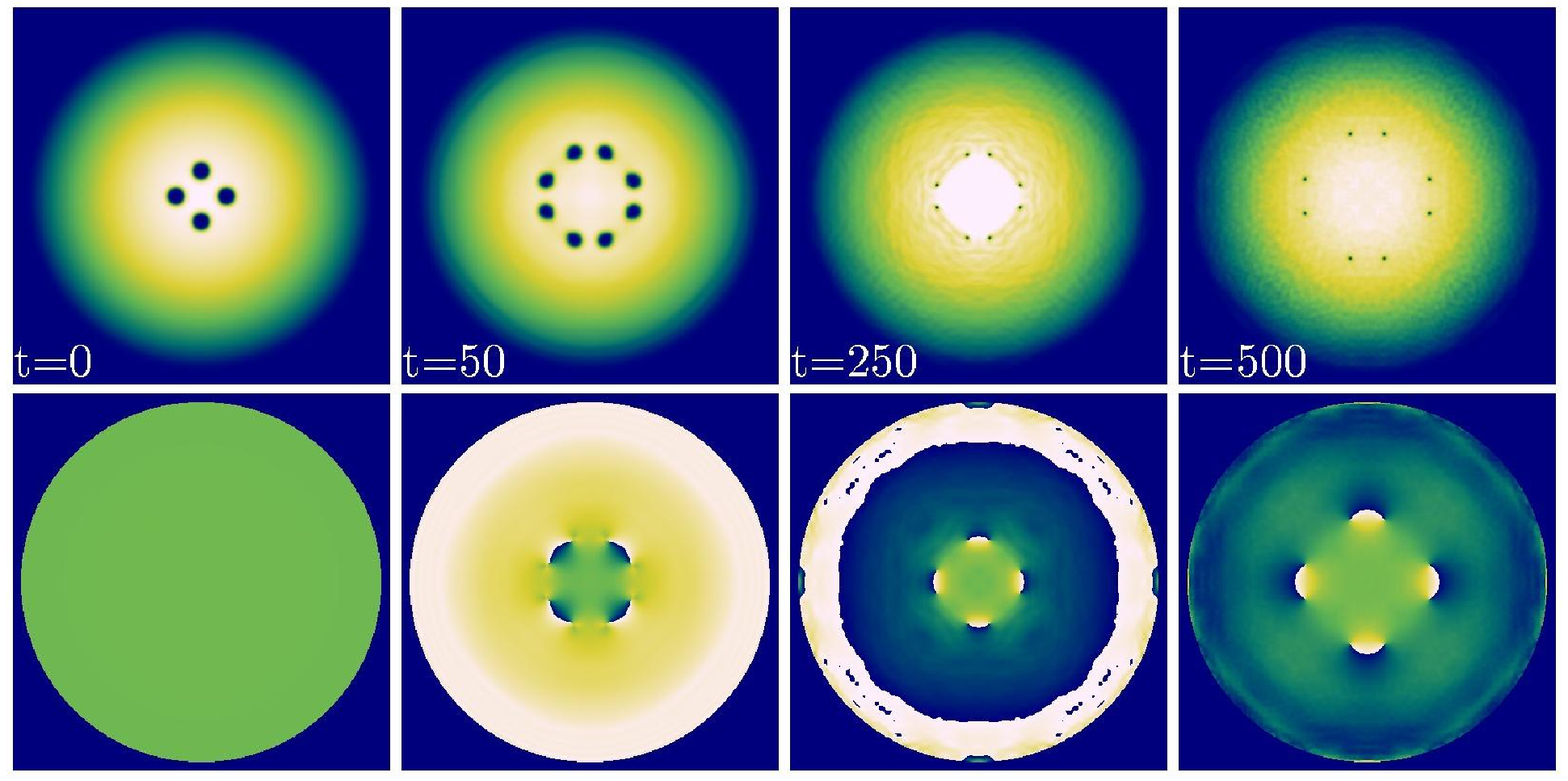}
\end{center}
\caption{(Color online)
Controlled generation of four, six, and eight vortices.
Chosen parameters are: chemical potential $\mu=2$, beam heights~$2\mu$,
and beam widths~$b_{w}=6$~$\mu$m.
Upper set of panels: Two pairs of beams initially located at $(0,\pm 8)$~$\mu$m are
moved outward with $v/c=0.123$ until $t=80$, then their positions are kept fixed.
From $t=80$ until $t=280$ the beams are linearly ramped down, the vortices are finally released and the time-evolution is monitored.
Middle panel: Similar but for 3 pairs of beams initially located symmetrically at a distance
of 10~$\mu$m from the origin with linear ramp down from $t=70$ until $t=270$.
Lower panel: Similar but for 4 pairs of beams initially symmetrically located at a distance
of 15~$\mu$m from the origin and moved at a speed $v/c=0.15$ and linearly ramped down
from $t=50$ until $t=250$.
}
\label{fig:Neven}
\end{figure}

Next, we investigate the possibilities to create larger, even, numbers of vortices in a neutral configurations with four, six, and eight vortices. The idea is to start with two, three or four pairs of overlapping chopstick beams close to the center of the BEC and
use the same methodology as above to create vortices from each beam pair
with the same protocol as described above. Doing so, each chopstick nucleates and
moves an independent vortex that might be placed in a desired
location. The beams then are kept fixed at their final
destination and subsequently linearly and adiabatically ramped down
to release the vortices that in turn are free to evolve without
the chopsticks being present.
Figure~\ref{fig:Neven} shows that this methodology is indeed feasible for controllably
creating, moving and releasing configurations bearing four, six, and eight vortices.
In principle, this method can be straightforwardly generalized to
even larger number of vortices as long as there is enough room
within the BEC to move the chopsticks beams.
As it can be observed from the last density distributions, the motion and
removal (ramping down) of multiple chopsticks has a significant perturbative,
effect on the background density.
In fact, if the disturbances to the condensate are considered to be too strong,
it would be advantageous to increase the value of the chemical potential in order
to diminish the role of interference of the resulting sound waves.
For this purpose, in our examples with multiple chopsticks we choose a slightly
larger value of the chemical potential (when compared to the previous
results with a single chopstick pair) of~$\mu=2$.
In view of the disturbances created by a larger number of beams, we found that
starting close to the center of the BEC and moving the beams outward is
advantageous in comparison to starting further outwards and moving the beams
inward.

\subsection{Non-neutral configurations}

We now consider creating an odd number of vortices with a charge imbalance of one, such that there is one more or one fewer positively charged vortex compared with negatively charged vortices.  To create such a non-neutral distribution in a controllable and repeatable manner we
start by creating a neutral configuration as discussed in the previous section and then take one vortex out of the condensate. We first demonstrate this procedure starting with a pair of oppositely charged vortices, removing one of the vortices, and leaving a single vortex that can then be re-positioned at will. The creation of a larger non-neutral configuration of vortices follows a similar principle.

\begin{figure}
\begin{center}
\includegraphics[width = 1.0\linewidth]{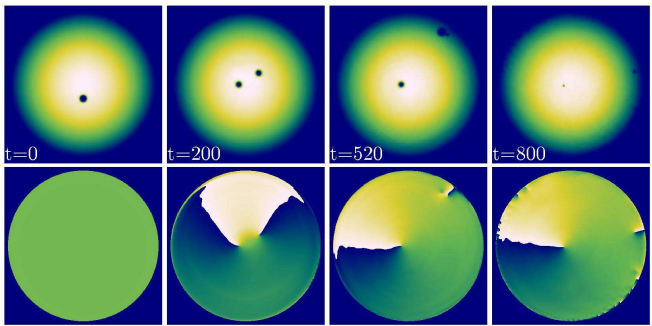}
\end{center}
\caption{(Color online) Controlled generation of a single vortex for $\mu=2$, beam height~$2\mu$ and beam width $b_{w}=6$~$\mu$m. The initial pair of beams located at $(0,-20)$~$\mu$m  is moved with $v/c=0.106$ to $(\pm 10,0)$~$\mu$m. One beam is then kept fixed at this position while the other one is moved further outwards, dragging the associated vortex with it.
At $t=500$ the former beam is ramped linearly down within~$\Delta t = 100$ to release
the single vortex close to the condensate center.
%
%
The chosen grid size for this case is $749\times 749$ resulting in a
spatial discretization with spacing~$dx=0.26$.}
\label{fig:N1}
\end{figure}

Figure~\ref{fig:N1} illustrates the ability to create a single vortex. First the protocol follows the generation of two vortices as in Fig.~\ref{fig:N2sym}. When the beams have reached $(\pm 10,0)$~$\mu$m, one of the beams is kept stationary at its position while the other beam is moved
further towards the edge of the condensate, dragging the trapped vortex with it.
Ideally, the vortex would be dragged all the way out of the
condensate. However, shortly before reaching the edge of the condensate the
vortex detaches from the beam and this edge vortex starts circulating
indefinitely around the condensate (see Fig.~\ref{fig:N1} where the vortex is
barely visible in the last two density snapshots, but is clearly visible in the
corresponding phase profile).
This type of ``detachment'' is a general issue that is encountered
when trying to ``rid'' the configuration of individual vortices by moving
one of the beams out of the BEC in a direction that is normal
to the edge of the BEC.
A number of pointers about how to bypass this issue, most notably
adjustments to the beam trajectory, and relying on the dissipation
stemming from thermal excitations, is discussed below.
As soon as the outgoing beam has reached the edge of the condensate the
first beam is linearly and adiabatically ramped down to release the remaining
single vortex to undergo free time evolution.

\begin{figure}
\begin{center}
\includegraphics[width = 1.0\linewidth]{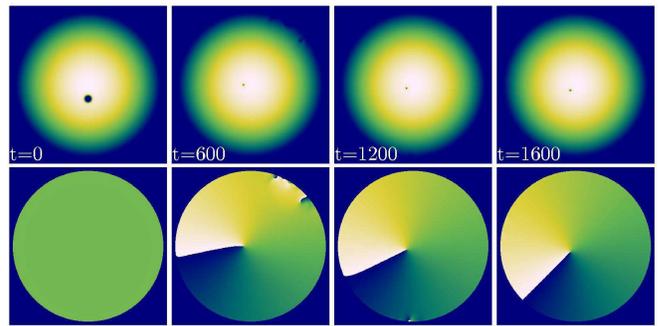}
\end{center}
\caption{(Color online) Similar to Fig.~\ref{fig:N1} but in the presence of dissipation,
as described by Eq.~(\ref{eq:diss}) with~$\gamma = 2\times 10^{-2}$.}
\label{fig:N1dissm2}
\end{figure}

\begin{figure}
\begin{center}
\includegraphics[width = 1.0\linewidth]{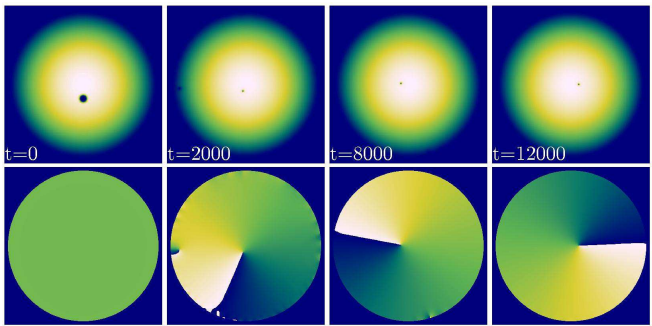}
\end{center}
\caption{(Color online) Similar to Fig.~\ref{fig:N1dissm2}
but in the presence of a more realistic dissipation value
of~$\gamma = 2\times 10^{-3}$
}
\label{fig:N1dissm3}
\end{figure}

While it is advantageous to reduce the beam velocity when dragging the vortex out
to allow the vortex to follow the beam trajectory closer to the edge of the condensate
(at the cost of larger time scales), it is ---typically, in our
observations--- not sufficient to drag the vortex all the way out of the BEC.
This is due to the fact that the density becomes small at the condensate edge
and thus the (density) contrast created by the laser beam close to this edge
is too small to pin or drag a vortex.
Nonetheless, we have found that the presence of dissipation (typically present
in all BEC experiments), as described by  Eq.~(\ref{eq:diss}), can resolve this issue.
Figure~\ref{fig:N1dissm2} depicts the results corresponding to Fig.~\ref{fig:N1}
with the addition of dissipation with $\gamma= 2\times 10^{-2}$.
As it is evident from the figure, at $t=1200$ the edge vortex is already very close
to the edge of the condensate (as is visible in the phase profile) and has completely
left the condensate by~$t=1600$.
Figure~\ref{fig:N1dissm3} shows a similar case but for the experimentally more realistic
value~$\gamma= 2\times 10^{-3}$; see the relevant discussion in Ref.~\cite{YanEtAl2014}, as
well as in earlier works on coherent structures such as dark solitons in Ref.~\cite{prouk23}.
In this case, the vortex still is visible at $t=8000$ and has completely left the
condensate at $t=12000$. We find the timescale for expelling the edge vortex from
the condensate ---with a starting point at about $t=600$ where the dragging beam has
reached the edge of the condensate--- to be approximately a factor ten times larger
than that for $\gamma= 2\times 10^{-2}$.
As the physical picture is qualitatively the same but the (computational) timescales
are considerable shorter in the following we will always opt to use the value
$\gamma= 2\times 10^{-2}$ to obtain results in the presence of dissipation.

\begin{figure}
\begin{center}
\includegraphics[width = 1.0\linewidth]{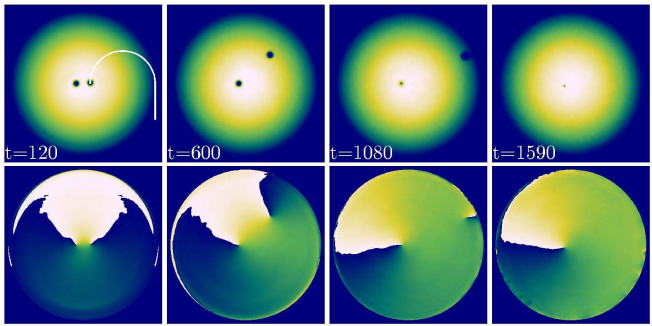}
\end{center}
\caption{(Color online)
Improved protocol to create a single vortex. Same as Fig.~\ref{fig:N1} but with a circular trajectory (as indicated by the white line in the left panel) instead of a straight line to drag the vortex out of the condensate. Additionally, the velocity of the dragging beam is reduced by 50\% after $t=110$. At~$t=1000$, when the latter beam starts to leave the condensate, the other beam is ramped down linearly within~$\Delta t=100$.}
\label{fig:Fig9}
\end{figure}

A more efficient protocol to drag a vortex out of the BEC relies on adjusting the laser
beam path such that its trajectory becomes more azimuthal as the beam gets closer
to the edge of the BEC and thus mimicking the natural tendency of the vortex
to precess about the trap center.
Figure~\ref{fig:Fig9} depicts this improved protocol where one of the beams
is moved on a circular path of diameter equal to the Thomas-Fermi radius
(see white line) such that it leaves the condensate tangentially.
Additionally, the velocity of the dragging beam is reduced by~50\% after $t=110$.
This results in the successful creation of a single vortex at~$t=1620$ as
the second vortex seems to be completely removed from the BEC
(it is neither visible in the density nor in the phase plot).

\begin{figure}
\begin{center}
\includegraphics[width = 1.0\linewidth]{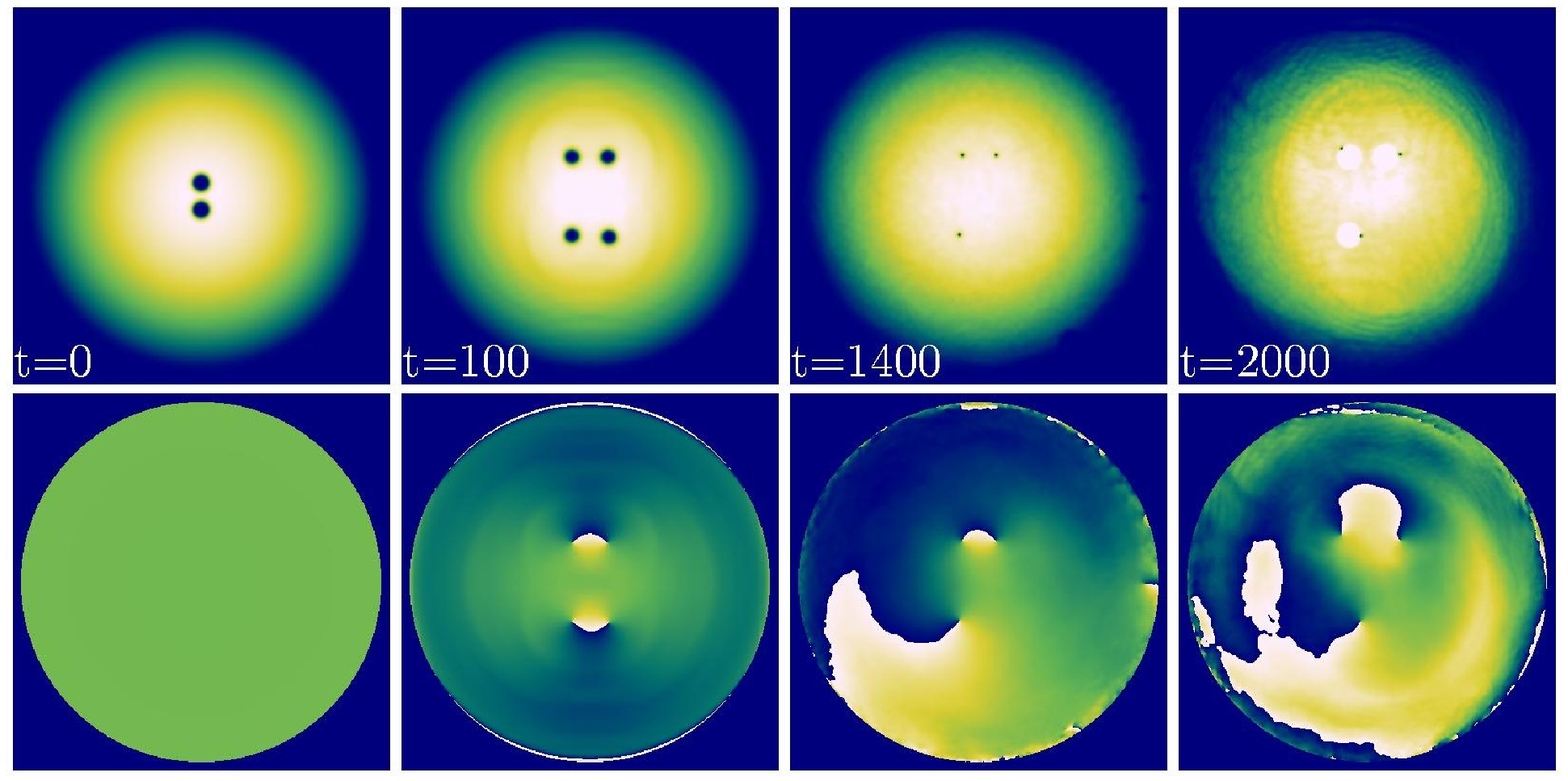}
\includegraphics[width = 1.0\linewidth]{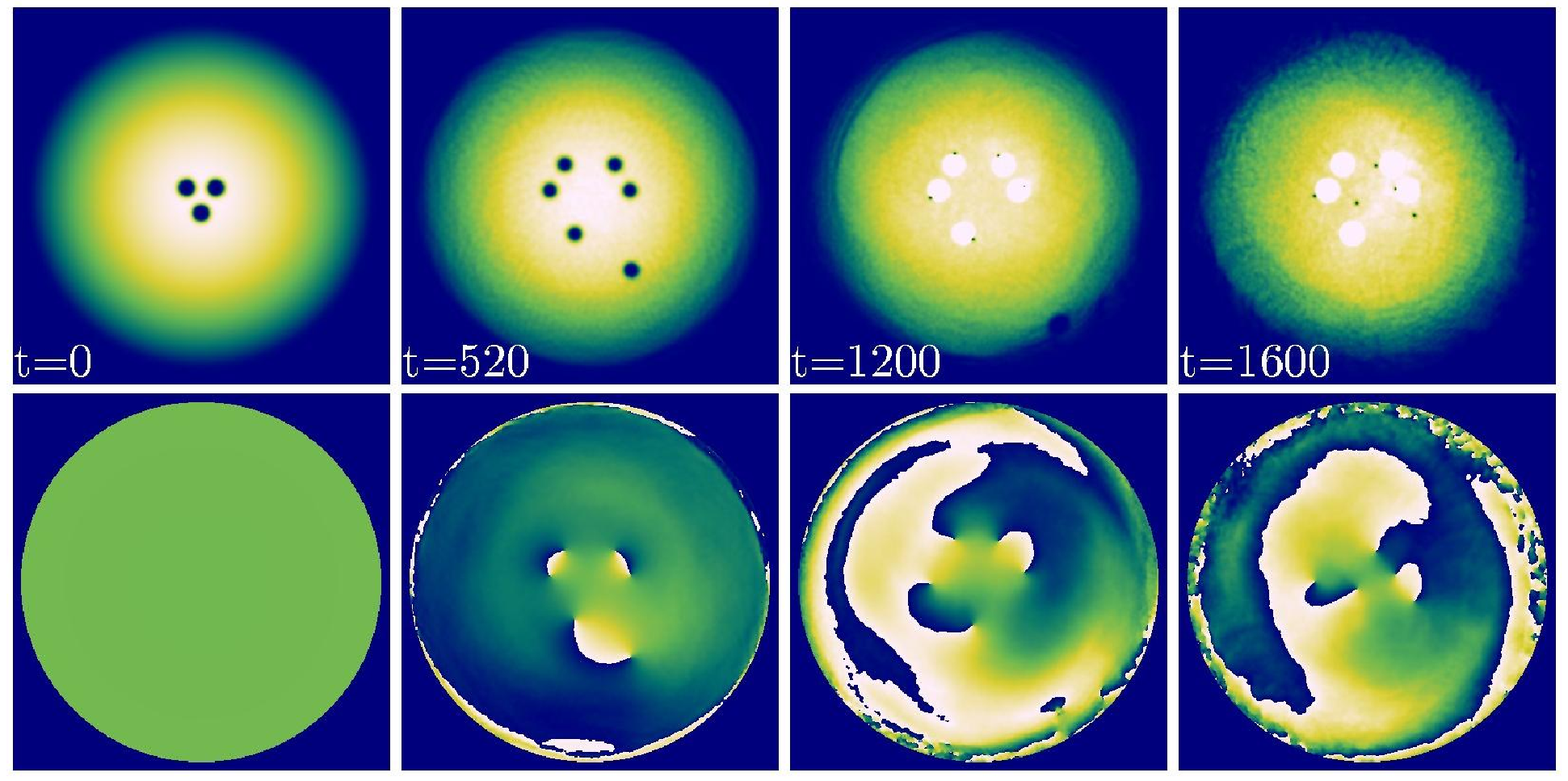}
\includegraphics[width = 1.0\linewidth]{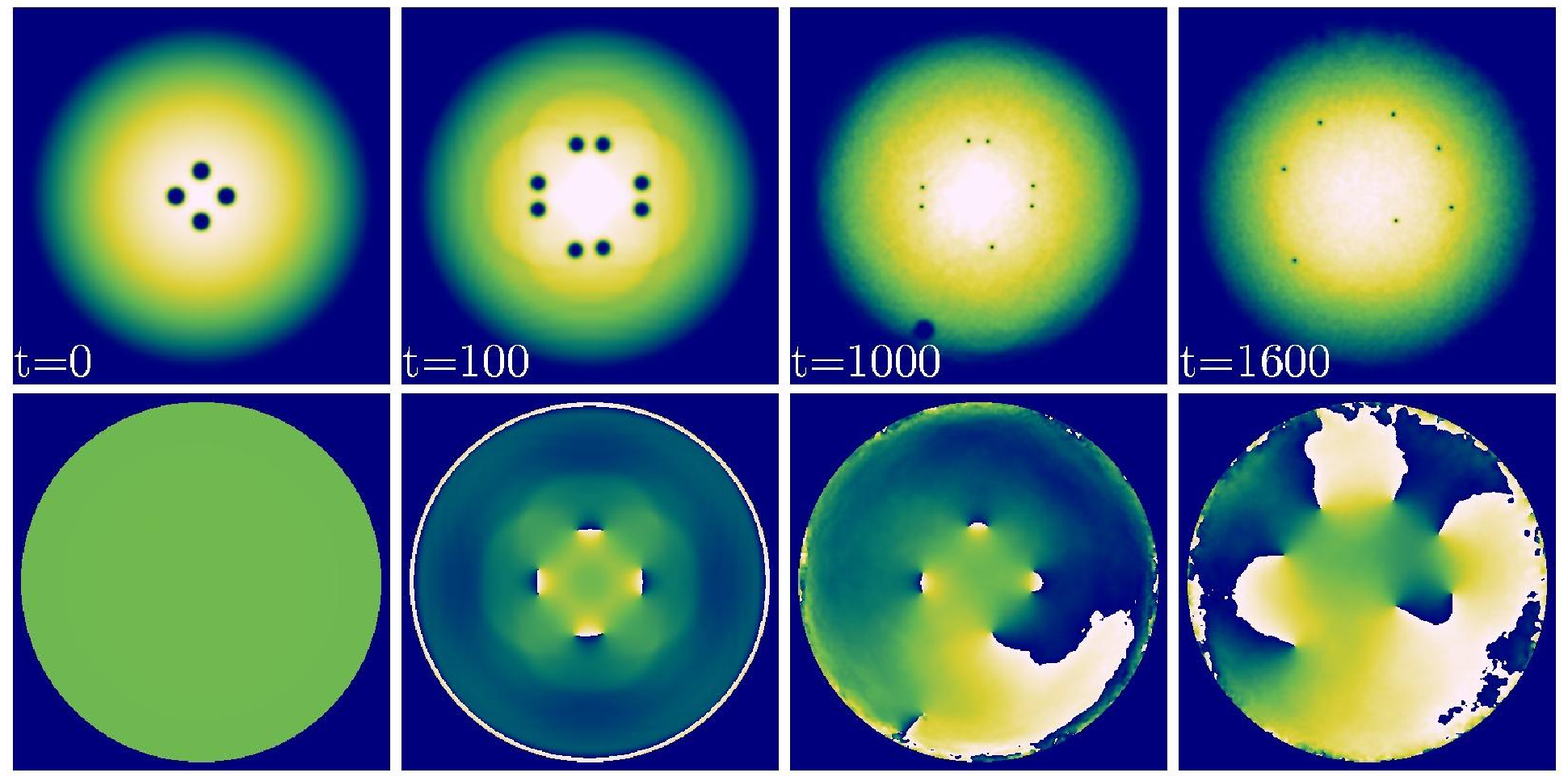}
\end{center}
\caption{(Color online) Controlled generation of an odd number of three to seven
vortices for the same parameters and initial configurations as in Fig.~\ref{fig:Neven}.
Upper rows: At $t=80$, three beams are kept fixed, while beam number four is moved
further outwards dragging one vortex out of the condensate. The beam velocity
is reduced to 0.25\% of its original value.
After linearly ramping down from $t=1200$ to $t=1400$ the other vortices
are released from their beams.
Middle rows: Similar but for six vortices stemming from three pairs of beams.
Five beams are kept fixed at $t=65$ and linearly ramped down from $t=600$ to $t=800$,
leading to a five vortex configuration, while the other is moved out of the condensate,
again with reduced speed of 0.25\% of the original value.
Bottom rows: Similar to the previous cases but for four pairs of beams.
Seven beams are kept fixed at $t=75$ and linearly ramped down from $t=800$ to $t=1000$
while one vortex is dragged all the way out of the condensate, endowing
the system with a configuration of seven vortices of alternating charge.}
\label{fig:Nodd}
\end{figure}

For the generation of three, five or seven vortices with a charge imbalance of one,
the idea is the same: start with a neutral configuration with for four, six or eight
vortices (see Fig.~\ref{fig:Neven}) and displace one beam outward with the aim of
removing a single vortex from the neutral configuration.
More specifically, after an even number of vortices is created, all but one
of the beams then are kept stationary at their respective locations
while the remaining beam is moved further outward to ideally drag the trapped vortex
all the way out of the condensate. This is followed by linearly ramping down the
other beams and releasing the vortex configuration in order
for it to perform free (i.e., unaffected by the beams beyond this time)
time evolution.
The results for this proposed methodology are displayed in Fig.~\ref{fig:Nodd}.
The case for initially four (and subsequently three) vortices is depicted in the
top two of panels in Fig.~\ref{fig:Nodd}. This case is more successful than for
the creation of a single vortex, i.e., here the remaining vortex has nearly
vanished in the background.
The case corresponding to initially six (and subsequently five) vortices
is depicted in the middle rows of panels in Fig.~\ref{fig:Nodd}.
%
%
In this case, the edge vortex is still visible at $t=1200$ but cannot be
distinguished from the background at $t=1600$.
Finally, the bottom rows of panels in Fig.~\ref{fig:Nodd} illustrate
the results for initially eight (and subsequently seven) vortices.
This case example is again fairly successful in the sense that a
potential edge vortex cannot be distinguished from the background.

\subsection{Comparison with vortex imprinting method}

The dynamics of vortices can also be explored by imprinting vortex solutions onto
the ground state BEC in the presence of only the harmonic confinement.
Such a technique has been used successfully in generating coherent
structures~\cite{sengstock}, although it should be noted that
typically in these experiments only a phase imprinting is induced.
The latter necessitates the ``morphing'' of the density around the
imprinting spot into a vortex profile, a process which, in turn, generates
considerable sound wave emission clearly visible in some of the case
examples of Refs.~\cite{sengstock}. Here, when we refer to imprinting, we more
accurately mean both phase and density engineering, or ``implanting''
a vortex in the system, namely
imprinting the phase, concurrently with modulating the density
in order to produce an ``as nearly exact as possible'' vortex waveform
in the system. While the latter scenario is quite idealized and
not straightforwardly achievable practically in the laboratory,
our aim is to compare the dynamics of our produced vortices via
the chopsticks method (containing the density modulations
induced by the light beams etc.) to ``target'' vortex dynamics
for a similar set of initial vortex locations.

\begin{figure}
\begin{center}
\includegraphics[width = 1.0\linewidth]{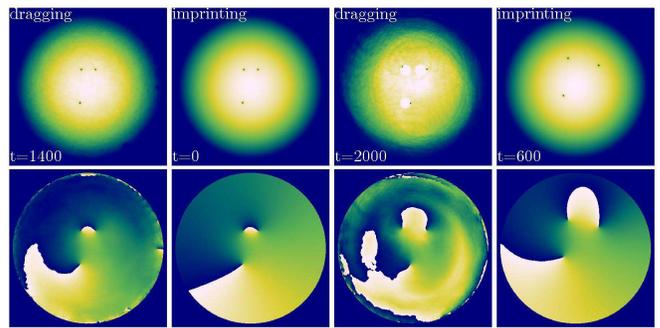}
\end{center}
\caption{(Color online) Comparison of the chopsticks protocol and the
vortex implanting method for the example of the three vortices from Fig.~\ref{fig:Nodd} for the configuration at $t=1400$ after releasing the vortices from the beams (first column). For the vortex implanting method the time evolution is initialized by a wave function with two negatively charged vortices of charge located at $(-10.025,-20.075)$ and $(9.25,21.5)$ and one positively charged vortex located at (-8.45,21.5) (second column).
Considerable differences in the time-evolution can be observed in comparison to the
time-evolution generated by the chopsticks method and the vortex implanting method's
dynamics (columns three and four, respectively)
due to the generation of sound waves in the former method.}
\label{fig:N3_imprinting}
\end{figure}

\begin{figure}
\begin{center}
\includegraphics[width = 1.0\linewidth]{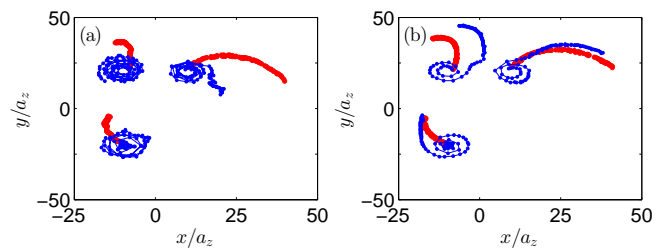}
\end{center}
\caption{(Color online)
Vortex trajectories corresponding to the cases depicted in
(a) Fig.~\ref{fig:N3_imprinting} and
(b) Fig.~\ref{fig:N3_imprinting_dgpe}.
The orbits in blue correspond to the chopsticks method while the
red orbits correspond to the implanting method.
}
\label{fig:Ntrajectories}
\end{figure}

In the following we compare the time evolution that is initialized by the
chopsticks method and compare it with the results obtained for the vortex
implanting method. It should be stressed that while the chopsticks method
generates (sometimes large) sound waves in the condensate as well
as allowing vortices to move prior to fully ramping off the laser beams,
the implanting method generates the most pristine setting with minimal sound
creation and nearly ``pure'' vortex dynamics.
Note that this is a numerical comparison only, as there have not been any demonstrations
of phase engineering and imprinting of arbitrary vortex distributions into a BEC.
We choose the case of three vortices, as displayed in the top rows of
Fig.~\ref{fig:N3_imprinting}. To initialize the vortex implanting method we
determine the positions and charge  of the three vortices (resulting from two
sets of split beams as explained in the previous subsection) at $t=1400$.
The ground state solution of the BEC is then multiplied with the
corresponding (normalized) vortex solutions (obtained for a
homogeneous background BEC) at the desired locations
(this is the ``implanting'' part). Fig.~\ref{fig:N3_imprinting}
clearly shows that the two methods generate a non-trivially
differing time evolution. While the qualitative structure of the
vortex positioning and trajectories when using the
chopsticks has been found to follow that of the ``implanted''
vortex configuration with the same initial vortex positions,
the density modulations induced by the
vortex generating beams definitely affect the precise aspects
of the vortex dynamics.
The comparison of the corresponding trajectories using both methods
is depicted in Fig.~\ref{fig:Ntrajectories}(a).

\begin{figure}
\begin{center}
\includegraphics[width = 1.0\linewidth]{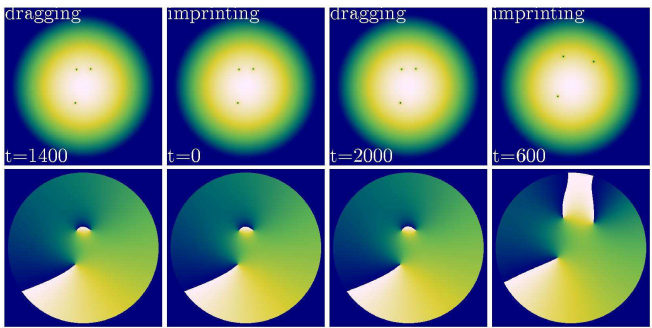}
\end{center}
\caption{(Color online) Same as Fig.~\ref{fig:N3_imprinting} but in the
presence of dissipation with $\gamma= 2\times 10^{-2}$.}
\label{fig:N3_imprinting_dgpe}
\end{figure}

Next, and for reasons of completeness,
we investigate the influence of dissipation on both cases.
While the presence of dissipation with $\gamma= 2\times 10^{-2}$ again
helps to expel the edge vortex from the condensate that is visible at~$t=1400$
(and barely visible at $t=2000$) for the vortex dragging method it does not
have a big influence on the motion of the other three vortices closer to the
BEC center. Also for the vortex implanting case the influence is negligible,
cf.~Fig.~\ref{fig:N3_imprinting_dgpe}. Hence, unfortunately, in this
case the presence of thermal excitations does not considerably
alleviate this discrepancy.
The comparison of the corresponding trajectories using both methods
is depicted in Fig.~\ref{fig:Ntrajectories}(b).

\subsection{Repositories}

Lastly, we explore the possibility of depositing several vortices in
so-called repository beams. The latter can be important for various
reasons in the form of persistent currents~\cite{philips,kody}, but also
towards the monitoring of
dynamics of large vortex clusters aggregating with a single sign
of vorticity, so-called Onsager-Kraichnan condensates,
also thermodynamically representing negative temperature states; see, e.g., the
recent analysis of Ref.~\cite{briannew}.

\begin{figure}
\begin{center}
\includegraphics[width = 1.0\linewidth]{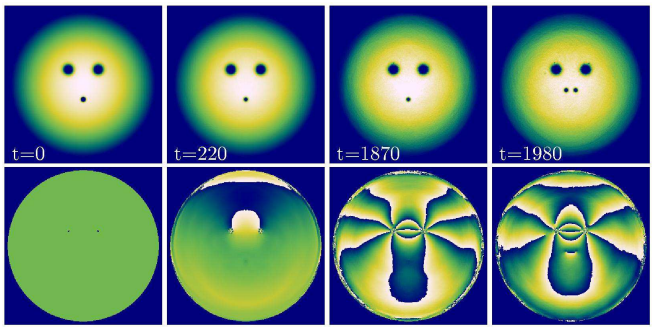}
\end{center}
\caption{(Color online) Principle of repository beams: Two stationary large beams with $b_{w}=15$~$\mu$m and beam height $2\mu$ serve as repository beams to collect vortices of equal charge. Vortices are created in oppositely charged pairs by ramping up two chopstick beams with $b_{w}=6$~$\mu$m at $(0,-30)$~$\mu$m (at $t=0$ they are already present) and by moving the beams to the pinning sites at $(\pm 30,30)$~$\mu$m. When the beams reach the pinning sites they are ramped down while new chopstick beams are ramped up at $(0,-30)$~$\mu$m. At $t=220$ one vortex is present in each repository beam, at $t=1870$ seven vortices are present in each pinning site. If a certain critical number of vortices per pinning site is exceeded vortices start leaking out of the repository beams, cf., panel for $t=1980$ where a vortex has been expelled from
the repository beam.
For this value of the chemical potential ($\mu=4$), the grid size is increased to
$1199\times 1199$, and~$dx=0.23$.
}
\label{fig:rep}
\end{figure}

As an example, we investigate a configuration where two stationary
repository beams with large beam waist are located in the BEC, cf.~Fig.~\ref{fig:rep}.
The idea is to create one pair of oppositely charged vortices after
another and deposit the positively charged vortices in one of the repositories
while the negatively charged vortices are deposited in the other repository
beam. In this way, persistent currents can be obtained~\cite{philips}.
The vortex generation portion of the sequence follows the principle illustrated in
Fig.~\ref{fig:N2sym}. When the chopsticks reach the repository beams
their position is kept fixed and they are linearly ramped down while a
new chopsticks pair is linearly ramped up and another pair of
vortices is created and dragged to the repository beams.
This procedure can
be repeated until a critical number of vortices is trapped in each repository
beam. As soon as a certain number of vortices ---depending on the size of the
repository beam--- is exceeded, vortices start leaking out of the
repositories. A natural constraint in that regard is that the repository
beam width should be larger than the product of the number of vortices
times their corresponding length scale, i.e., the healing length of
the BEC.

To create a large number of vortices within this procedure with as little disturbance of the condensate as possible, it is advantageous to use a larger chemical potential. In the following, we choose~$\mu=4$. The size of the chopsticks beams is again~$b_{w} = 6$~$\mu$m. In Fig.~\ref{fig:rep} we display results for repository beams with~$b_{w} = 15$~$\mu$m. The time evolution is initialized by
using the stationary state with two chopsticks beams present at $(0,-30)$~$\mu$m and repository beams located at $(\pm 30,30)$~$\mu$m. For a duration~$\Delta t=175$ the chopstick beams are moved towards $(\pm 30,30)$~$\mu$m, creating and dragging the first pair of vortices with them. At the position of the repository beams the chopstick beams are linearly ramped down within~$\Delta t = 100$ and the respective vortices are consequently deposited in the repository beam. While the old chopstick beams are ramped down
a new chopstick pair is linearly ramped up at the same initial position as the former,
again within~$\Delta t = 100$. Then the whole procedure is repeated and the second vortex dipole pair is created, dragged along and deposited,
respectively,
within the repository beams. For the chosen size of the repository beams seven vortices can be placed into each beam. After this critical number has been reached, vortices start leaking out of the repository beams. At~$t=1980$ the first vortex has been expelled from the repository beam.

\begin{figure}
\begin{center}
\includegraphics[width = 1.0\linewidth]{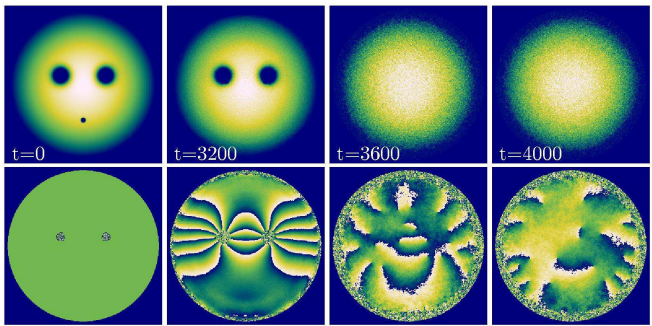}
\end{center}
\caption{(Color online)
Similar as Fig.~\ref{fig:rep} but with larger chemical potential~$\mu=5$ and larger repository beams with~$b_{w}=30$~$\mu$m.
In this way ten vortices can be trapped per repository beam. Within $\Delta t=233$ the dragging beams are moved toward the repositories and ramped up/down within $\Delta t = 100$. At $t=3230$, the repository beams are linearly ramped down within $\Delta t = 100$ to release the vortices to the free time evolution.
However, importantly a considerable amount of sound emission is present
in the dynamics due to the significant density modulations imposed
by the repository beams.
For this value of the chemical potential ($\mu=5$) the grid size is further
increased to $1499\times 1499$, resulting in~$dx=0.21$.
}
\label{fig:rep20}
\end{figure}

\begin{figure}
\begin{center}
\includegraphics[width = 1.0\linewidth]{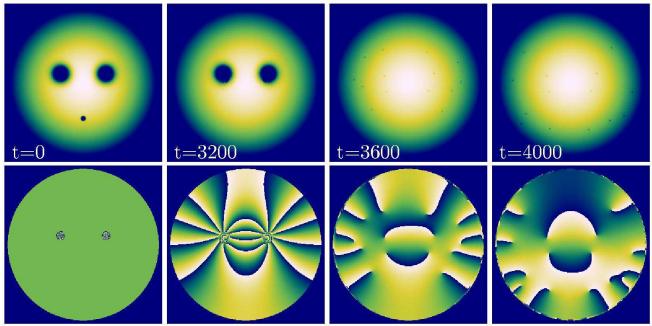}
\end{center}
\caption{(Color online) Same as Fig.~\ref{fig:rep20} but in the presence of
dissipation with~$\gamma=2\times 10^{-2}$. Notice that the sound waves
have been chiefly expelled from the BEC, enabling a more pristine
monitoring of the vortex dynamics.
}
\label{fig:rep20dGPE}
\end{figure}

To accommodate even more vortices within the repositories, we increase the chemical
potential to $\mu=5$ and the size of the repository beams to $30$~$\mu$m.
For these parameters ten vortices can be deposited in each repository,
cf.~Fig.~\ref{fig:rep20}. Next, both the chopstick and the repository beams
are linearly ramped down within $\Delta t=100$ to release the vortices to
their free time evolution.
Figure~\ref{fig:rep20dGPE} shows the same protocol as in Fig.~\ref{fig:rep20} but
in the presence of dissipation with~$\gamma=2\times 10^{-2}$.
It can be seen that the introduction of dissipation is, in fact, beneficial: the significant
disturbances (relevant sound waves) to the atomic cloud due to the density modulations induced
by the repositories, that can be seen in Fig.~\ref{fig:rep20} are considerably smoothed
out in the dissipative case.

This specific configuration with two repository beams can easily be modified to other interesting configurations. Note that a repository beam sitting on the edge of the condensate can serve as a location to trap edge vortices that can arise when a vortex is supposed to be taken out of the condensate as in Fig.~\ref{fig:N1} in a reproducible manner. However, note that
our results demonstrate that the presence of dissipation helps to get rid of the edge vortices while the vortices closer to the center are less affected by the latter.

\section{Conclusions \& Future Challenges}
\label{sec:conclu}

In the present work, we have numerically explored a mechanism that enables
the production and manipulation of multiple quantized vortices, essentially at will, inside
an atomic Bose-Einstein condensate. The use of lasers as vortex ``optical tweezers''
in a judicious manner, so as to gradually create the phase profile associated with a pair of oppositely charged vortices, as well as to pin vorticity and
create persistent current, enabled us
to locate vortices at various positions within the BEC by moving the laser beams.
We were subsequently able to transfer these vortices at will, including
moving them outside the BEC or positioning them within a
repository beam. Repeating the process either with multiple
optical beams or after the delivery of the first pair to the repository beams,
starting the process anew, we were able to produce arbitrary neutral
and non-neutral vortex distributions. Naturally, the process has a number
of limitations, such as the emergence of density modulations due to the carrying
beams (which also to some extent affect the vortex motions),
or in some cases the difficulty of carrying individual vortices
outside the condensate (due to the so-called ``detachment'' from
the beam in regions of low density near the condensate rims.
With respect to most of these aspects, the contribution of thermal
fluctuations in the context of the so-called dissipative
Gross-Pitaevskii equation is beneficial, enabling the outward
motion of vortices and also the partial reduction of sound waves.
The method is also generally limited by the number of laser beams
of finite width that can be located and moved within a BEC of finite radius.

This technique creates a broad set of possibilities that are quite worthwhile
for subsequent experimental and further numerical exploration. As far as we
know, no other technique available in the literature is as versatile towards
creating/engineering multiple vortices while
selecting their charge and position distributions at will.
Constructing and understanding the dynamics of such vortices and vortex
clusters~\cite{middel11,Ree2014.PRA89.053631}, especially in the context
of anisotropy where they can be robust, e.g., in collinear
states~\cite{jan11,recentjpa} and exploring more systematically their
stability is now a tractable and experimentally realizable topic.
Further studies oriented towards devising methods of decreasing the generation
or effects of residual sound waves so as to enable a more direct engineering of
free-vortex states and initialization and study of vortex dynamics would certainly
be desirable.
Finally, extending these types of methods to higher dimensional BECs in three
dimensions in the case of vortex lines
and vortex rings would be of particular interest in its own right.

\begin{acknowledgments}
We would like to thank Q.-Y.~Chen for discussions and numerical
assistance during the early stages of this project.
B.G.~acknowledges support from the European Union through FP7-PEOPLE-2013-IRSES Grant Number 605096.
P.G.K.~acknowledges support from
the National Science Foundation under grant DMS-1312856,
from the European Union through FP7-PEOPLE-2013-IRSES Grant Number 605096, and
from the Binational (US-Israel) Science Foundation through grant 2010239.
R.C.G.~acknowledges support from DMS-1309035.
B.P.A.~is supported by the National Science Foundation under grant PHY-1205713.
P.G.K.'s work at Los Alamos is supported in part by the U.S. Department of Energy.
The computations were performed on the HPC cluster HERO, located at the
University of Oldenburg and funded by the DFG through its Major Research
Instrumentation Programme (INST 184/108-1 FUGG), and by the Ministry of
Science and Culture (MWK) of the Lower Saxony State.
\end{acknowledgments}

\bibliographystyle{unsrt}

\end{document}